\documentclass[useAMS,usenatbib]{mn2e}
\usepackage{graphicx}
\usepackage{epstopdf}

\begin{document}

\title[Redshift evolution as a test of dark energy]{The time evolution of cosmological redshift as a test of dark energy}

\author[A. Balbi and C. Quercellini]{A. Balbi $^{1,2}$\thanks{E-mail: balbi@roma2.infn.it} and C. Quercellini $^{1}$\thanks{E-mail: claudia.quercellini@uniroma2.it}\\
$^{1}$Dipartimento di Fisica, Universit\`a di Roma ``Tor Vergata'',
via della Ricerca Scientifica 1, 00133 Roma, Italy\\
$^{2}$INFN Sezione di Roma ``Tor Vergata'',
via della Ricerca Scientifica 1, 00133 Roma, Italy} 

\date{\today}

\pagerange{\pageref{firstpage}--\pageref{lastpage}} \pubyear{2007}

\maketitle

\label{firstpage}

\begin{abstract}
The variation of the expansion rate of the Universe with time produces an evolution in the cosmological redshift of distant sources (for example quasar Lyman-$\alpha$ absorption lines), that might be directly observed by future ultra stable, high-resolution spectrographs (such as CODEX) coupled to extremely large telescopes (such as European Southern Observatory's Extremely Large Telescope, ELT). This would open a new window to explore the physical mechanism responsible for the current acceleration of the Universe. We investigate the evolution of cosmological redshift from a variety of dark energy models, and compare it with simulated data. We perform a Fisher matrix analysis and discuss the prospects for constraining the parameters of these models and for discriminating among competing candidates. We find that, because of parameter degeneracies, and of the inherent technical difficulties involved in this kind of observations, the uncertainties on parameter reconstruction can be rather large unless strong external priors are assumed. However, the method could be a valuable complementary cosmological tool, and give important insights on the dynamics of dark energy, not obtainable using other probes.
\end{abstract}


\section{Introduction}
The discovery of the current accelerated expansion of the Universe \citep{Riess:astro-ph/9805201, Perlmutter:astro-ph/9812133} has prompted a renewed interest towards classic cosmological tests. The measurement of the expansion rate of the Universe at different redshifts is crucial to investigate the cause of the accelerated expansion, and to discriminate candidate models. Until now, a number of cosmological tools have been successfully used to probe the expansion and the geometry of the Universe. The position of acoustic peaks in the cosmic microwave background (CMB) angular power spectrum provides a powerful geometrical test, showing that the space curvature of the Friedmann-Robertson-Walker (FRW) metric is nearly flat \citep{Spergel:astro-ph/0603449}. A similar test is performed through the detection of baryon acoustic oscillations (BAO) in the power spectrum of matter extracted from galaxy catalogues. The luminosity distance of type Ia supernovae and other standard candles allows to constrain the value of the expansion rate at different redshifts, typically up to $z\sim2$ \citep{Riess:astro-ph/0611572}. 

Currently, however, very little is known about the detailed dynamics of the expansion. Depending on the underlying cosmological model, one expects the redshift of any given object to exhibit a specific variation in time. An interesting issue, then, is to study whether the observation of this variation, performed over a given time interval, could provide useful information on the physical mechanism responsible for the acceleration, and be able to constrain specific models. This is the main goal of this paper. In addition to being a direct probe of the dynamics of the expansion, the method has the advantage of not relying on a determination of the absolute luminosity of the observed source, but only on the identification of stable spectral lines, thus reducing the uncertainties from systematic or evolutionary effects. The possible application of this kind of observation as a cosmological tool was first pointed out by \cite{1962ApJ...136..319S}. However, the tininess of the expected variation (typically, a shift of less than a cm/s over a period of a year) was deemed impossible to observe at the time. The importance of this test was stressed again over the past decades by other authors (e.g.\ \citealt{1981ApJ...247...17L}, \citealt{1980ApJ...240..384R}); more recently, \cite{1998ApJ...499L.111L} re-assessed its feasibility and concluded that, given the advancement in technology occurred over the last forty years, it is conceivable to expect that a measurement of the redshift variation in the spectra of some suitable source (most notably the Lyman-$\alpha$ absorption lines of distant QSOs) could be detected in the not too distant future. Recently, \cite{Lake:astro-ph/0703810v1} showed that measuring the time evolution of redshift would be a way to check the internal consistency of the underlying cosmological model, and to map the equation of state of dark energy. 

With the foreseen development of extremely large observatories, such as the European ELT, with diameters in the range 30-100 m, and the availability of ultra-stable, high-resolution spectrographs, the perspectives for the observation of redshift variations look very promising. For example, \cite{2005Msngr.122...10P, 2006IAUS..232..193P} pointed out that the CODEX (COsmic Dynamics Experiment) spectrograph should have the right accuracy to detect the expected signal by monitoring the shift of Lyman-$\alpha$ lines of distant ($z\ga 2$) QSOs over a period of some decades. 

A previous investigation of the expected cosmological constraints from this kind of observations was performed by \cite{2007PhRvD..75f2001C}. That work, however, only made predictions for a very restrictive set of models and assumptions: firstly, it only explored the case when the dark energy is a standard cosmological constant (i.e.\ a component with constant equation of state $w=p/\rho=-1$) plus a few non-standard models (the Chaplygin gas, and a model that, although named interacting dark energy, in fact only generalises a constant dark matter equation of state, leaving the cosmological constant untouched); secondly, as it will be shown in more detail later, the observational strategy considered in that paper seems rather optimistic.

In this paper, we aim to give an overview of the theoretical predictions for the most popular and still viable models introduced in the recent past to explain the observed accelerated expansion. These models either fall within the broad category of ``dark energy'', introducing an unknown, smooth and gravitationally repulsive component, or invoke a modification of the theory of gravity (for a review of possible explanation for the accelerated expansion see, e.g., \cite{2003RvMP...75..559P}). Our first goal is to investigate whether the dynamics of any of these standard and non-standard dark energy models shows interesting features that could be constrained by future observations of the redshift variation. We do not restrict ourselves {\it a priori} to models with flat geometry, and we fully take into account correlations between parameters when assessing the expected uncertainties. 

This paper is organised as follows. In Sect. 2 we review the basic equations that describe the redshift variation with time of a source, in an expanding universe. We comment on the possibility of detecting the effect in the near future in Sect. 3. In Sect. 4 we study the expected apparent velocity shift for a number of non-standard dark energy models. Finally, we discuss the main results and conclusions of our work in Sect. 5 and 6. 

\section{Basic theory}
We start by reviewing the basic theory necessary to derive the expected redshift variation in a given cosmological model. We assume that the metric of the Universe is described by the FRW metric. The observed redshift of a given source, which emitted its light at a time $t_s$, is, today (i.e. at time $t_0$), 
\begin{equation}
z_s(t_0)=\frac{a(t_0)}{a(t_s)}-1,
\end{equation}
and it becomes, after a time interval $\Delta t_0$ ($\Delta t_s$ for the source)
\begin{equation}
z_s(t_0+\Delta t_0)=\frac{a(t_0+\Delta t_0)}{a(t_s+\Delta t_s)}-1.
\end{equation}
The observed redshift variation of the source is, then, 
\begin{equation}
\Delta z_s=\frac{a(t_0+\Delta t_0)}{a(t_s+\Delta t_s)}-\frac{a(t_0)}{a(t_s)},
\end{equation}
which can be re-expressed, after an expansion at first order in $\Delta t/t$, as:
\begin{equation}
\Delta z_s\simeq\Delta t_0\left(\frac{\dot a(t_0)-\dot a(t_s)}{a(t_s)}\right).
\end{equation}
Clearly, the observable $\Delta z_s$ is directly related to a change in the expansion rate during the evolution of the Universe, i.e.\ to its acceleration or deceleration, and it is then a direct probe of the dynamics of the expansion. It vanishes if the Universe is coasting during a given time interval (i.e.\ neither accelerating nor decelerating). We can rewrite the last expression in terms of the Hubble parameter $H(z)=\dot a(z)/a(z)$:
\begin{equation}
\label{deltaz}
\Delta z_s=H_0\Delta t_0\left(1+z_s-\frac{H(z_s)}{H_0}\right).
\end{equation}
The function $H(z)$ contains all the details of the cosmological model under investigation. Finally, the redshift variation can also be expressed in terms of an apparent velocity shift of the source, $\Delta v=c\Delta z_s/(1+z_s)$.

\section{Can the velocity shift be observed?}

The latest studies on the feasibility of detecting a time evolution of the redshift are those by \cite{2005Msngr.122...10P, 2006IAUS..232..193P}. The most promising approach is to look at the spectra of Lyman-$\alpha$ forest absorption lines, which are very stable and basically immune from peculiar motions. This is a scenario that might be achieved in the next decades, when extremely large telescope (such as the ELT) will collect a large number of photons, and high-resolution spectrographs (such as CODEX) will be able to measure tiny shifts in spectral lines over a reasonable amount of time, typically of order of few decades.

According to Monte Carlo simulations discussed by \cite{2005Msngr.122...10P, 2006IAUS..232..193P}, the accuracy of the spectroscopic velocity shift measurements expected by CODEX can be modelled as:
\begin{equation}\label{forecast}
\sigma_{\Delta v}=1.4\left( \frac{2350}{S/N} \right)\left( \frac{30}{N_{QSO}} \right)^{1/2}\left( \frac{5}{1+z_{QSO}} \right)^{1.8} \mathrm{cm/s};
\end{equation}
here, $S/N$ is the signal to noise ratio for pixels of 0.0125 \AA, $N_{QSO}$ is the number of QSOs spectra observed and $z_{QSO}$ is their redshift.
Based on the currently known QSOs brighter than $m\leq 16.5$ in the redshift range $2\la z \la 4$, \cite{2005Msngr.122...10P, 2006IAUS..232..193P} assumed to observe either 40 QSOs with $S/N$ ratio of 2000 each, or 30 QSOs with $S/N$ of 3000, respectively. In fact, again according to \cite{2005Msngr.122...10P}, a CODEX-like experiment, coupled to a 60 m telescope with approximately 20$\%$ total efficiency, would give a cumulative $S/N$ of 12000 for a single QSO, requiring roughly 125 hours of observation to get a $S/N$ of 3000 on that spectrum. Then, starting with 10 hours of observation per night, and taking into account a 20$\%$ use of the telescope, and a $90\%$ of actually usable data, one finds that 40 spectra can be measured with that $S/N$ in roughly 7.6 years (this time would actually increase to about 15 years if the telescope aperture is smaller, e.g., 42 m instead of 60 m). Then, it seems that a reasonable time span to perform a second observation of the same spectra might be 30 years. 

Based on the previous considerations, our study will then be conducted assuming a future dataset containing a total of 40 QSOs spectra (uniformly distributed over 5 equally spaced redshift bins in the redshift range 2-5), with a $S/N=3000$, observed twice over a time interval of 30 years.   We note that a previous study performed by \cite{2007PhRvD..75f2001C} seems to make optimistic claims, i.e. that 240 pairs QSO's spectra can be observed with $S/N=3000$ over a time span of 10 years. It is easily shown (with the same arguments as above) that with these figures the required observing time would actually be roughly 90 years for each of the two epochs. Moreover, even assuming an increase of the number of QSOs in the redshift range 2-4 with future large all-sky photometric surveys, it seems quite difficult to predict an order of magnitude increase over the current known objects, which are 25 \citep{2005Msngr.122...10P}.

Using the expected error bars from Eq. \ref{forecast}, we can predict the level of accuracy that can be reached in the reconstruction of the parameters for a comprehensive set of dark energy models. Furthermore, we can predict whether any of these models can be distinguished from the standard $\Lambda$CDM scenario. We will now look into these problems.

\section{Predictions for dark energy models}
All of the models we will consider in this paper are currently viable candidates to explain the observed acceleration, i.e.\ they have not been falsified by available tests of the background cosmology. Clearly, some models may be preferred with respect to others based on some statistical assessment of their ``economy'', i.e.\ the fact that they fit the data well with a smaller number of parameters.  Given the current status of cosmological observations, there is no strong reason to go beyond the simple, standard cosmological model with zero curvature and a cosmological constant $\Lambda$ (except for the conceptual problems arising when one attempts to reconcile its observed value with some estimate derived from fundamental arguments, see, e.g., \citealt{Weinberg:1988cp}).  For the scope of the present paper, however, it is interesting to explore as many models as possible, since future observations of the time variation of redshift could reach a level of accuracy which could allow to better discriminate competing candidates, and to understand the physical mechanism driving the expansion. We refer the interested reader to the paper by \cite{Davis:astro-ph/0701510}, which discusses the constraints on most of the models that we will focus on in our study. Unless stated otherwise, throughout our paper we assume for each class of models the best fit values found in that work, and vary the parameters within their 2$\sigma$ uncertainties. 

All the predictions on the time evolution of redshift presented in the following were derived assuming observations performed over a time interval $\Delta t_0=30$ years. From Eq.~\ref{deltaz} it is clear that the expected velocity shift signal increases linearly with $\Delta t_0$, so that it is straightforward to calculate the expected signal when a different period of observation is assumed. Fig. \ref{fig:all} shows our predictions for the cosmological models discussed in the following, along with simulated data points and error bars derived from Eq. \ref{forecast}. 

\subsection{The standard cosmological model ($\Lambda$CDM)}\label{lambda}

We start our analysis by first setting out the predictions for the current standard cosmological model. In the simplest scenario, the dark energy is simply a cosmological constant, $\Lambda$, i.e.\ a component with constant equation of state $w=p/\rho=-1$. Flatness of the FRW metric is usually assumed, but in general one can parametrize the deviation from the zero-curvature case in terms of the parameter $\Omega_k\equiv1-\Omega$ where $\Omega$ is the total density of the Universe in units of the critical value $\rho_c=3H_0^2/8\pi G$.

The Hubble parameter evolves according to the Friedmann equation, which, for this model, is:
\begin{equation}
\left(\frac{H}{H_0}\right)^2=\frac{\Omega_k}{a^2} + \frac{\Omega_m}{a^3} +  \Omega_\Lambda,
\end{equation}
where $\Omega_m$ and $\Omega_\Lambda$ parameterize the density of matter and cosmological constant, respectively. When flatness is assumed, $\Omega=\Omega_m+\Omega_\Lambda=1$, and the model has only one free parameter, $\Omega_\Lambda$. The current best fit value from cosmological observations is $\Omega_\Lambda=0.73\pm 0.04$ in the flat case \citep{Davis:astro-ph/0701510}, while relaxing the assumption of flatness results in a preference for slightly closed models, with $\Omega_k=-0.01\pm 0.01$ \citep{Spergel:astro-ph/0603449}.

\subsection{Dark energy with constant equation of state}\label{constantw}
The next step is to allow for deviations from the simple $w=-1$ case, introducing a component with an arbitrary, constant value for the equation of state. The accelerated expansion is obtained when $w< -1/3$. The Hubble parameter for this generic dark energy component with density $\Omega_{de}$ then becomes:
\begin{equation}
\left(\frac{H}{H_0}\right)^2= \frac{\Omega_k}{a^2} + \frac{\Omega_m}{a^3} + \frac{\Omega_{de}}{a^{-3(1+w)}} .
\end{equation}
The currently preferred values of $w$ in this models still include the cosmological constant case: $w=-1.01\pm 0.15$ \citep{Davis:astro-ph/0701510}.

\subsection{Dark energy with variable equation of state}\label{linder}

If the equation of state of dark energy is allowed to vary with time, then the Hubble parameter is:
\begin{equation}
\left(\frac{H}{H_0}\right)^2=\frac{\Omega_k}{a^2} + \frac{\Omega_m}{a^3} + \frac{\Omega_{de}}{e^{3\int da(1+w(a))/a}}.
\end{equation}
In this case, one has to choose a suitable functional form for $w(a)$, which in general involves a parametrization. The most commonly used \citep{2001IJMPD..10..213C,2003PhRvL..90i1301L} is:
\begin{equation}\label{eq:linder}
w(a)=w_0+w_a(1-a),
\end{equation}
although different approaches can be used. Clearly, the exact form of $w(a)$ with time will lead to completely different evolution for the dark energy component.

\subsection{Interacting dark energy}\label{coupled}

In interacting dark energy models the dark components interact through an energy exchange term. The conservation equations for matter and dark energy can be written in a very general way as
\begin{eqnarray}
&&\dot{\rho_m}+3H\rho_m=\delta H \rho_m,\\
&&\dot{\rho_{de}}+3H\rho_{de}(1+w)=-\delta H \rho_m,
\end{eqnarray}
so that the total energy-momentum tensor is conserved.
Whenever $\delta$ is a non-zero function of the scale factor, the interaction causes $\rho_m$ and $\rho_{de}$ to deviate from the standard scaling, and the mass of the particles is not conserved.
This non-standard behaviour has been parametrized \citep{2001PhRvL..87n1302D,2004astro.ph.10543M} by the relation $\rho_{de}/\rho_m=Aa^\xi$, where $A=\Omega_{de}/(1-\Omega_{de}-\Omega_k)$ and the density parameters are the present quantities. The Hubble parameter or this model then reads
\begin{equation}
\left(\frac{H}{H_0}\right)^2=\frac{\Omega_k}{a^2}+a^{-3}\left(1-\Omega_k\right)\left(1-\Omega_{de}\left(1-a^\xi\right)\right)^{-3\frac{w}{\xi}},
\end{equation}
which reduces to the uncoupled case for $\xi=-3w$. This model also includes all late-time scaling solutions. We also note that this model is a genuinely interacting dark energy, unlike the one discussed in \cite{2007PhRvD..75f2001C} which is actually a generalised dark matter (thus more similar to the model we discuss in \ref{affine}).

\subsection{DGP models}\label{dgp}

The Dvali-Gabadadze-Porrati (DGP) model \citep{2000PhLB..485..208D} provides a mechanism for accelerated expansion which is alternative to the common repulsive-gravity fluid approach: within the context of brane-world scenarios, the leaking of gravity in the bulk, above a certain cosmologically relevant physical scale, is responsible for the increase in the expansion rate with time. The only parameter of this class of models is $r$, the length at which the leaking occurs, which defines an adimensional parameter $\Omega_{r}\equiv 1/(4r^2H_0^2)$. The Hubble parameter then reads:
\begin{equation}
\left(\frac{H}{H_0}\right)^2= \frac{\Omega_k}{a^2} + \left(\sqrt{\frac{\Omega_m}{a^3} + \Omega_{r}} + \sqrt{\Omega_{r}}\right)^2,
\end{equation}
where $\Omega_m=1-\Omega_k-2\sqrt{\Omega_{r}}\sqrt{1-\Omega_k}$.

\subsection{Cardassian models}\label{cardassian}

Another possibility originating from the brane-world scenario is that of a so-called Cardassian expansion \citep{2002PhLB..540....1F} resulting from a modification of the Friedmann equation, with the introduction of a term that depends non linearly on the average density of the Universe, assumed to be composed only of matter. This additional term, phenomenologically, is equivalent to the introduction of a dark energy component, with a scaling law $\propto a^{-3n}$, where $n$ is completely equivalent to the quantity $w+1$ of the dark energy models with constant equation of state. More interesting are the so-called ``modified polytropic Cardassian'' models, which have an extra parameter $q$ and a Friedmann equation:

\begin{equation}
\left(\frac{H}{H_0}\right)^2=\frac{\Omega_m}{a^3}\left( 1 +  \frac{\left(\Omega_m^{-q}-1\right)}{a^{3q(n-1)}} \right)^{1/q}.
\end{equation}

\subsection{Chaplygin gas}\label{chaplygin}
There are a few models which attempts to explain both structure formation and the current acceleration of the Universe with a single ``dark fluid'', whereas the standard scenario relies on two separate dark contributions to the stress-energy tensor (a dark matter and a dark energy component). A well-studied case is the so called Chaplygin gas \citep{2001PhLB..511..265K}, 
where the unified dark component has equation of state $p=\tilde{A}\rho^{-\gamma}$ with $\tilde{A}>0$. 

The expansion rate in this model is governed by the equation: 
\begin{equation}
\left(\frac{H}{H_0}\right)^2=\frac{\Omega_k}{a^2}+(1-\Omega_k)\left(A + \frac{\left(1-A\right)}{a^{3(1+\gamma)}} \right)^{1/(1+\gamma)}.
\end{equation}

with the definition $A=\tilde{A}/\rho_0^{1+\gamma}$, where $\rho_0$ is the total density of the Universe at the present. The so-called ``standard case'' for the Chaplygin gas is obtained for the choice $\gamma=1$ (which, however, is not a good fit to current data), while for $\gamma=0$ the model recovers the standard cosmological constant case with $\Omega_m=(1-\Omega_k)(1-A)$.

\subsection{Affine equation of state}\label{affine}
An interesting possibility to consider is that the dark energy is modelled by a generic, barotropic equation of state $p=p(\rho)$, as discussed in \cite{Chiba:astro-ph/9704199,Visser:gr-qc/0309109,Ananda:astro-ph/0512224}. In particular, the case where the Taylor expansion of an arbitrary equation of state of that sort is truncated to first order, e.g.\ $p=p_0+\alpha\rho$, has recently been investigated by \cite{2007astro.ph..2423B}, who also derived cosmological constraints on the parameters of the model. It is interesting to note that such an affine equation of state can be used to describe a simple unified dark matter model, with a time evolution of the background density given by $\rho(a)=\rho_\Lambda+(\rho_{o}-\rho_\Lambda)a^{-3(1+\alpha)}$, where $\rho_\Lambda\equiv -p_0/(1+\alpha)$ and $\rho_0$ is the dark energy fluid energy density at present. The Hubble parameter is given by:
\begin{equation}
\left(\frac{H}{H_0}\right)^2=\frac{\Omega_k}{a^2} + \frac{\tilde\Omega_m}{a^{3(1+\alpha)}} +  \Omega_\Lambda,
\end{equation}
where $\tilde\Omega_m\equiv (\rho_{o}-\rho_\Lambda)/\rho_c$; for $\alpha=0$ this model recovers the standard $\Lambda$CDM case. 

\section{Results and discussion}

We used the expected error bars from Eq.~\ref{forecast} to perform a Fisher matrix analysis \citep{1997ApJ...480...22T}, leading to an estimate of the best possible constraints on the parameters of dark energy models. All our predictions are based on 30 years of observation, assuming that the fiducial values for the parameters of each dark energy model are those which best fit current observations (as from \citealt{Davis:astro-ph/0701510}, \citealt{2007astro.ph..2423B}, \citealt{2004astro.ph.10543M}). 

The Fisher matrix formalism allows one to estimate the best possible accuracy attainable on the determination of the parameters of a certain model. Specifically, given a set of cosmological parameters $p_i$, $i=1,...,n$, and the corresponding Fisher matrix $F_{ij}$ (that is easily calculated based on a theoretical fiducial model and the assumed data errors from Eq. \ref{forecast}), the best possible $1\sigma$ error on $p_i$ is given by $\Delta p_i\equiv C_{ii}^{1/2}$, where the covariance matrix $C_{ij}$ is simply the inverse Fisher matrix: $C_{ij}=F_{ij}^{-1}$. It is a well-known fact (see, e.g.\ \citealt{1997MNRAS.291L..33B}) that, when estimating the expected errors $\Delta p_i$ on each parameters, one has to be careful about existing correlations with the other parameters. During the inversion process, non-null off-diagonal Fisher matrix elements propagate in the diagonal elements of the covariance matrix, giving their contribution to the estimated uncertainties. Neglecting this contribution, for example by simply taking the inverse of the elements of the Fisher matrix (a process which is equivalent to assume perfect knowledge of all the other parameters) usually results in severely underestimated errors. The existence of correlations among estimated parameters is a manifestation of degeneracies: when more and more parameters are allowed to vary, and they have similar effects on the observable quantities, it becomes increasingly difficult to constrain each parameter. 

From the above considerations, two crucial consequences arise: first, dark energy models with intrinsically less free parameters will have an advantage with respect to models with more free parameters; second, not taking properly into account the degeneracies among parameters (for example, by assuming perfect knowledge on some of them) will lead to wrong estimates of the errors. For this reason, we allowed all the parameters which are specific of a given dark energy model to vary simultaneously in our analysis (i.e. we inverted the full Fisher matrix when estimating errors). However, since it would be pointless not to assume {\it any} prior knowledge, we fixed the parameters not directly related to dark energy (such as, for example, the baryon density) to their current best fit value.

As an example of what we just discussed, let us consider first the standard $\Lambda$CDM case. Assuming that the fiducial model has  $\Omega_\Lambda=0.7$ and $\Omega_k=0$ and that both $\Omega_\Lambda$ and $\Omega_k$ can vary,  we find $\Delta\Omega_\Lambda=0.2$ and $\Delta\Omega_k=0.25$ at 1$\sigma$.  When we fix $\Omega_k=0$, the bound on $\Omega_\Lambda$ becomes 0.007 (at 1$\sigma$). If dark energy is modelled by a constant equation of state (with a fiducial value $w=-1$) and the flatness constraint is imposed we find a looser bound on the dark energy density, $\Delta\Omega_{de}=0.016$, and quite a large error on the equation of state, $\Delta w=0.58$. This clearly shows that different assumptions on the knowledge of any parameter has an influence on all the others. Having noted this, however, we also note that the parameter $\Omega_k$ can be much better constrained using external datasets, such as the CMB anisotropy. For the sake of simplicity, then, we will assume flatness in the following.

The prospect of detecting departures from the standard  $\Lambda$CDM case could in principle be one of the real assets of observing the time evolution of redshift, and is thus worth closer investigation. Since the simulated data used in our analysis assume that QSOs are used as a tracer of the redshift evolution, we expect that the more constrained models will be those that have the largest variability in the redshift range $2\la z \la 5$. Of course, additional information could be obtained using other observational windows. We will comment on this issue later on. 

\begin{figure*}
\centering
   \includegraphics[width=7cm]{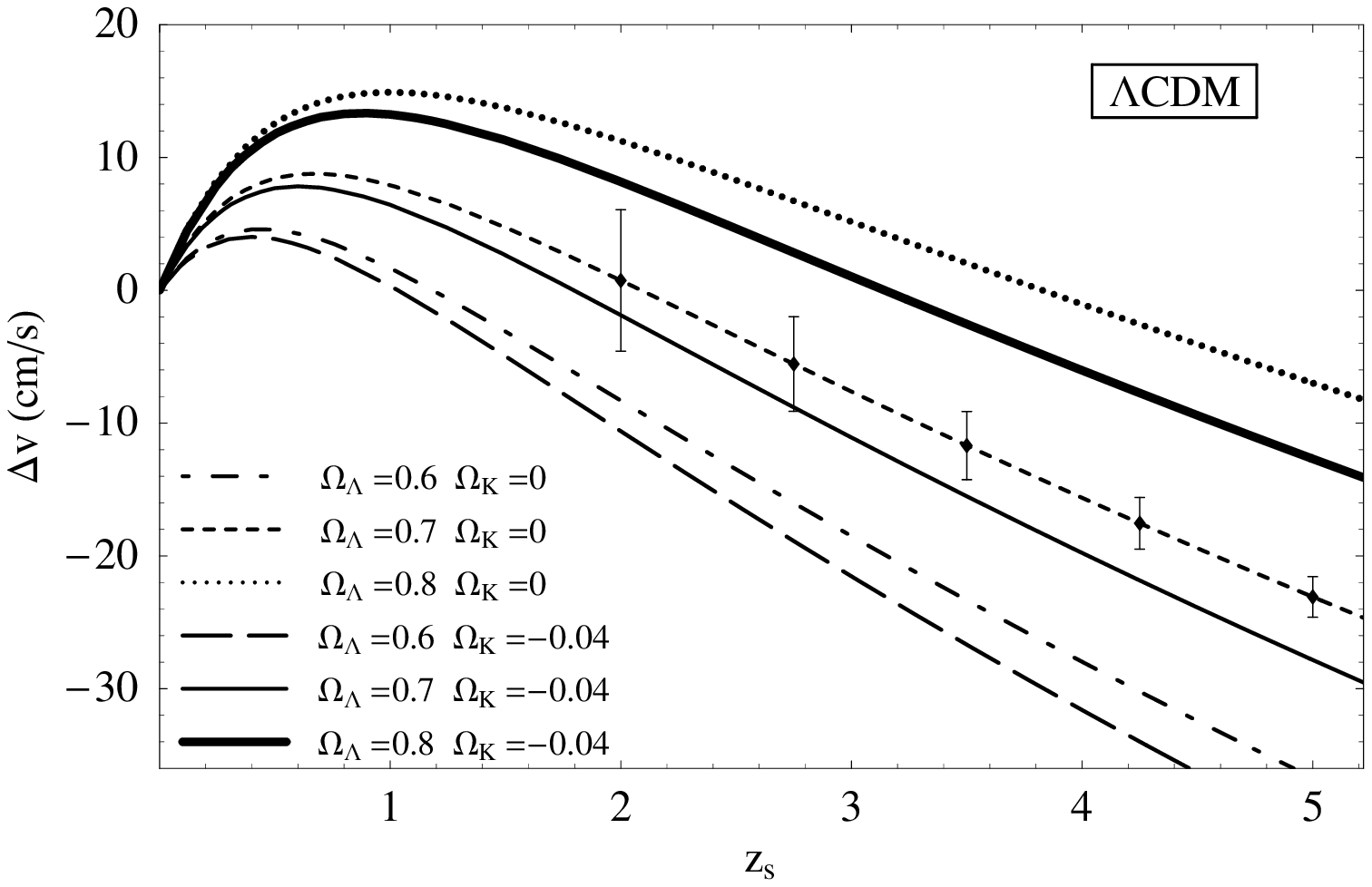}%
   \includegraphics[width=7cm]{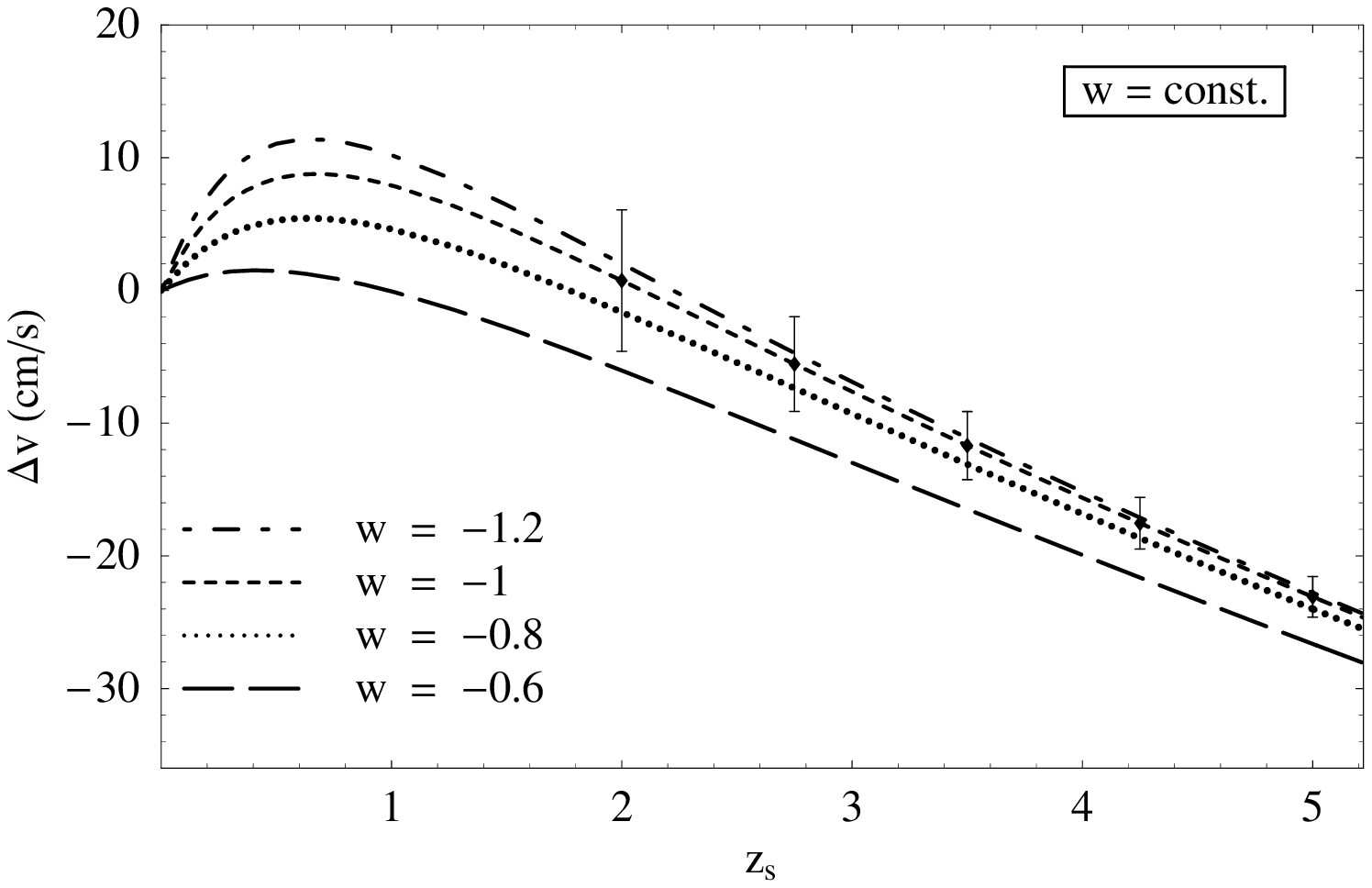}\\
   \includegraphics[width=7cm]{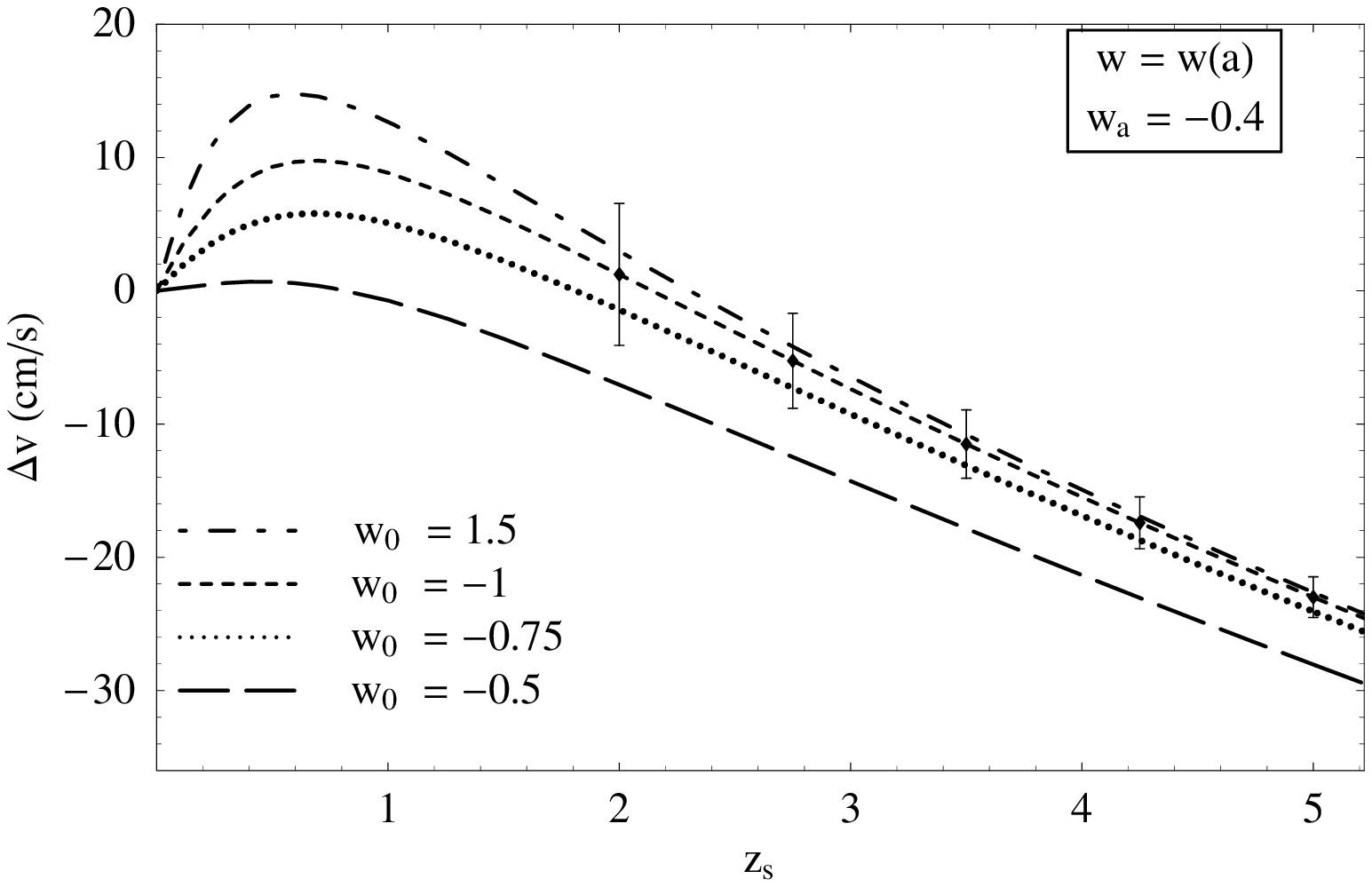}%
   \includegraphics[width=7cm]{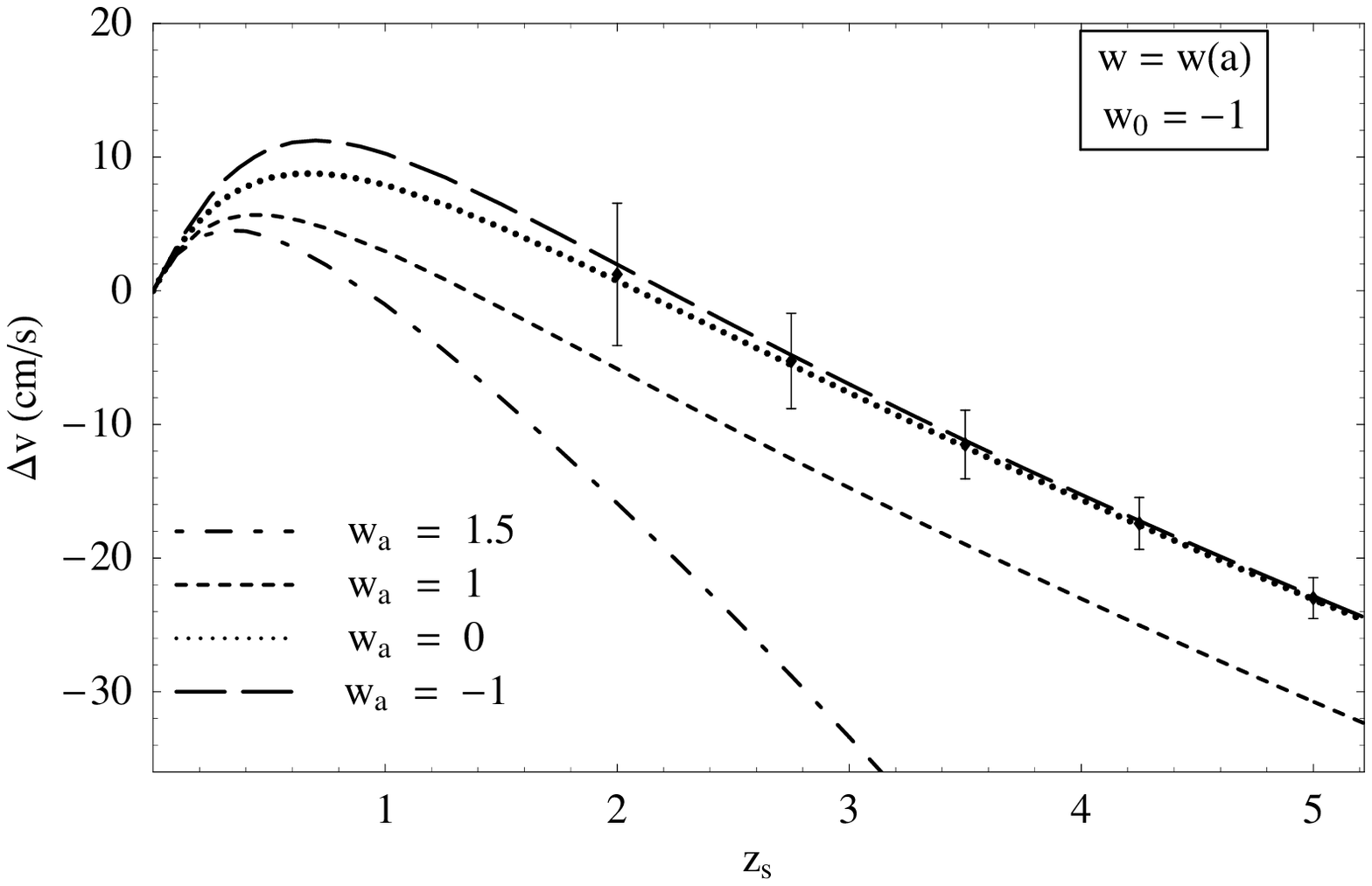}\\ 
   \includegraphics[width=7cm]{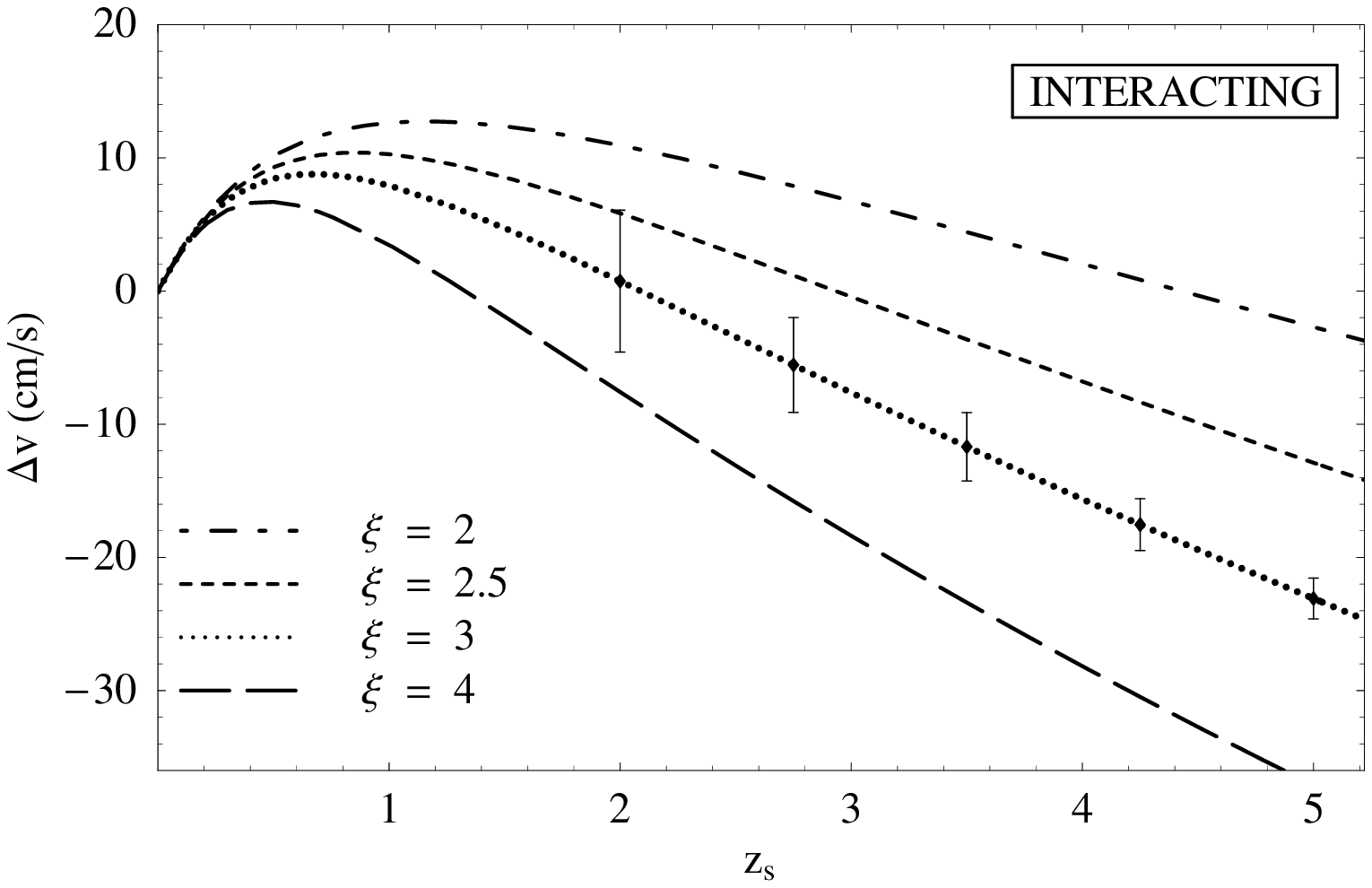}%
   \includegraphics[width=7cm]{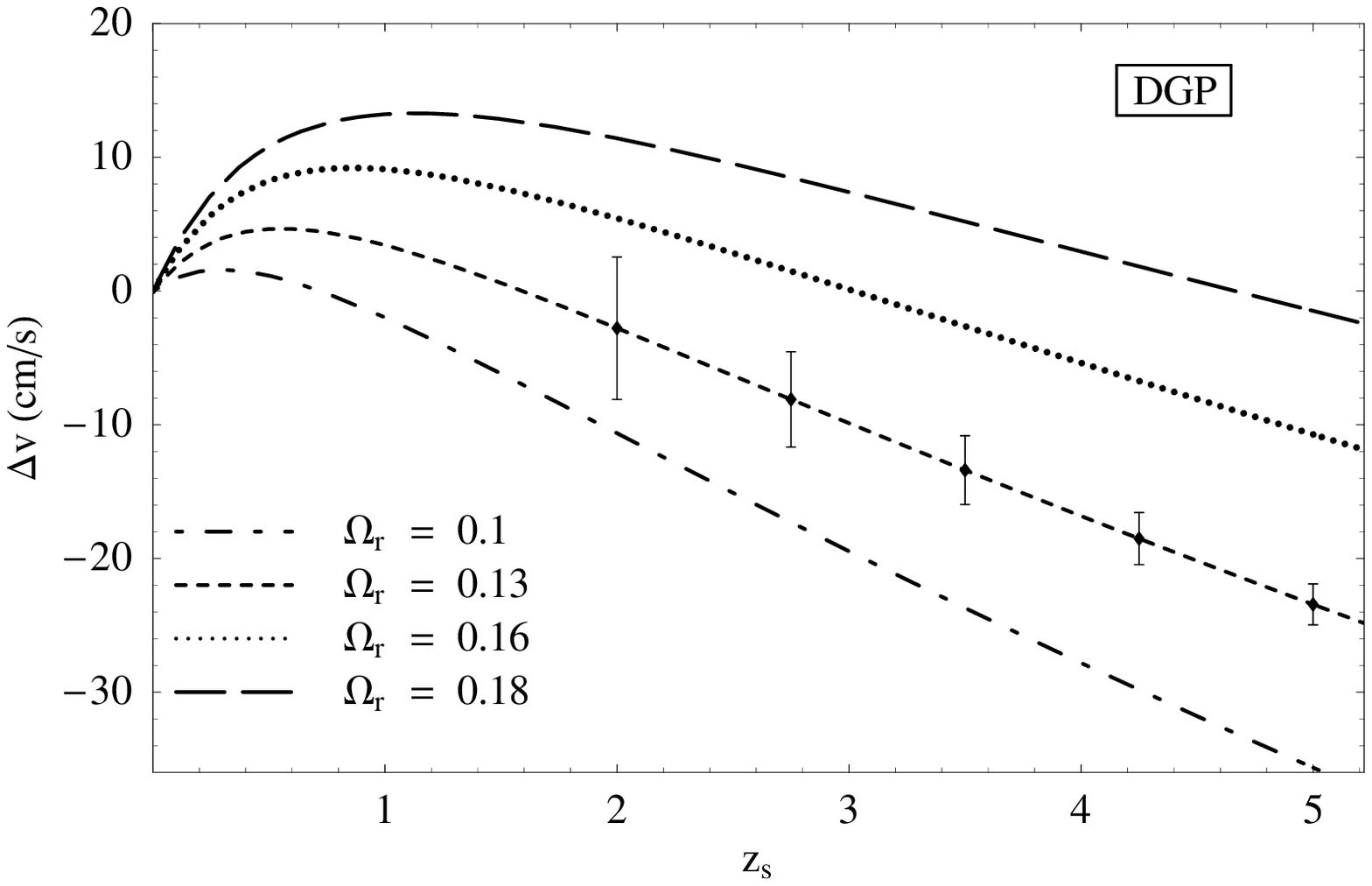}\\   
   \includegraphics[width=7cm]{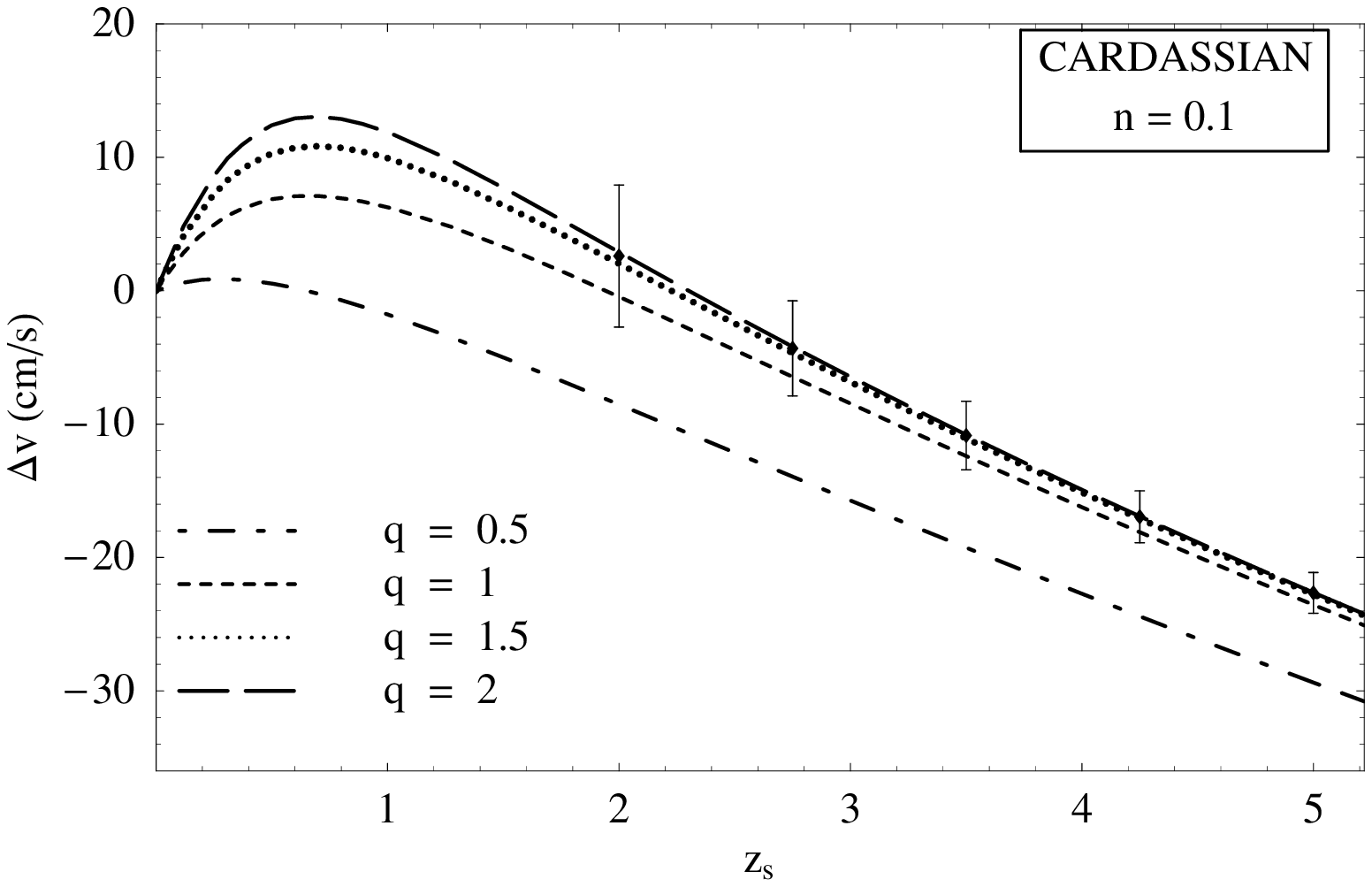}%
   \includegraphics[width=7cm]{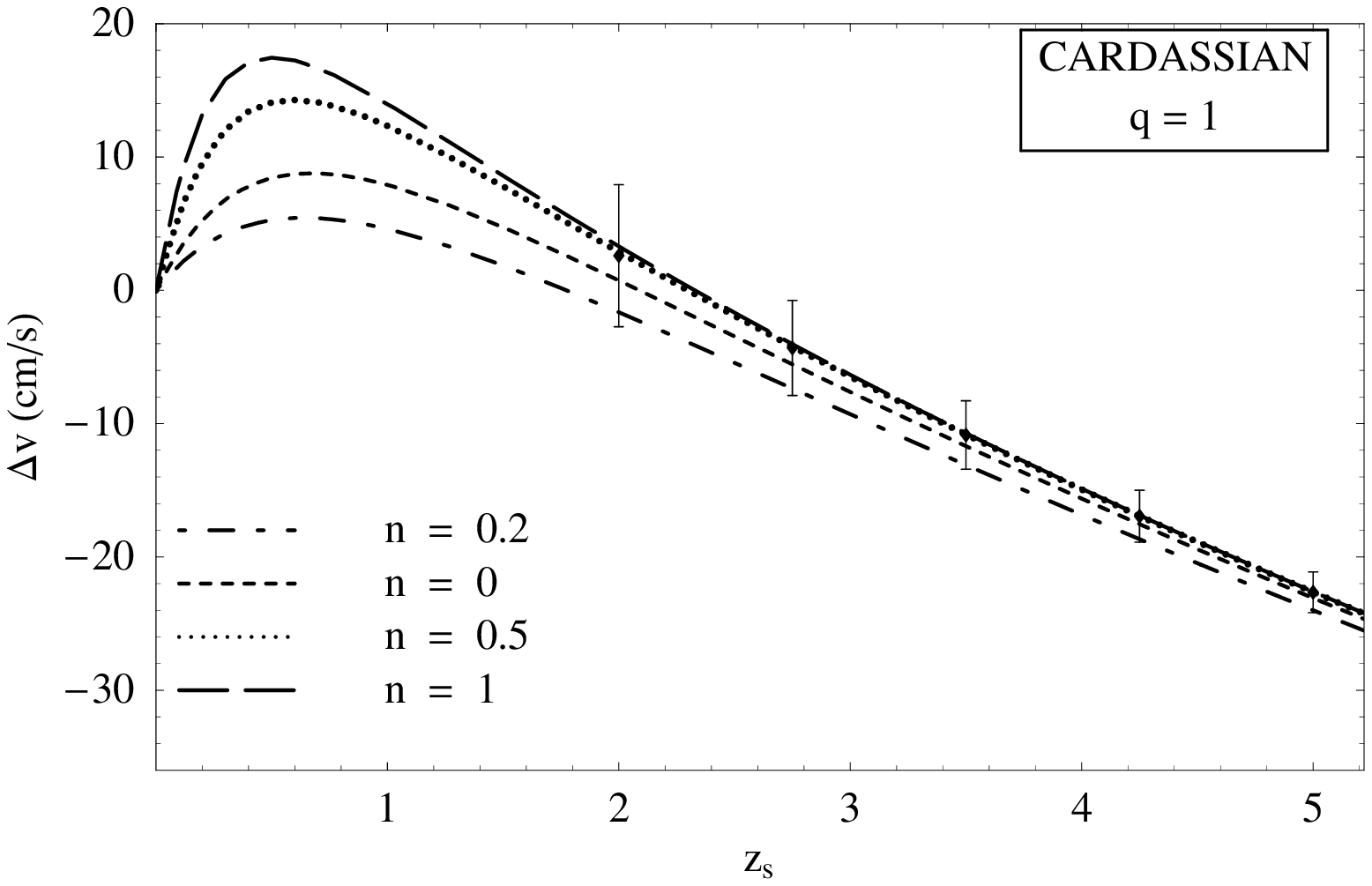}\\  
   \includegraphics[width=7cm]{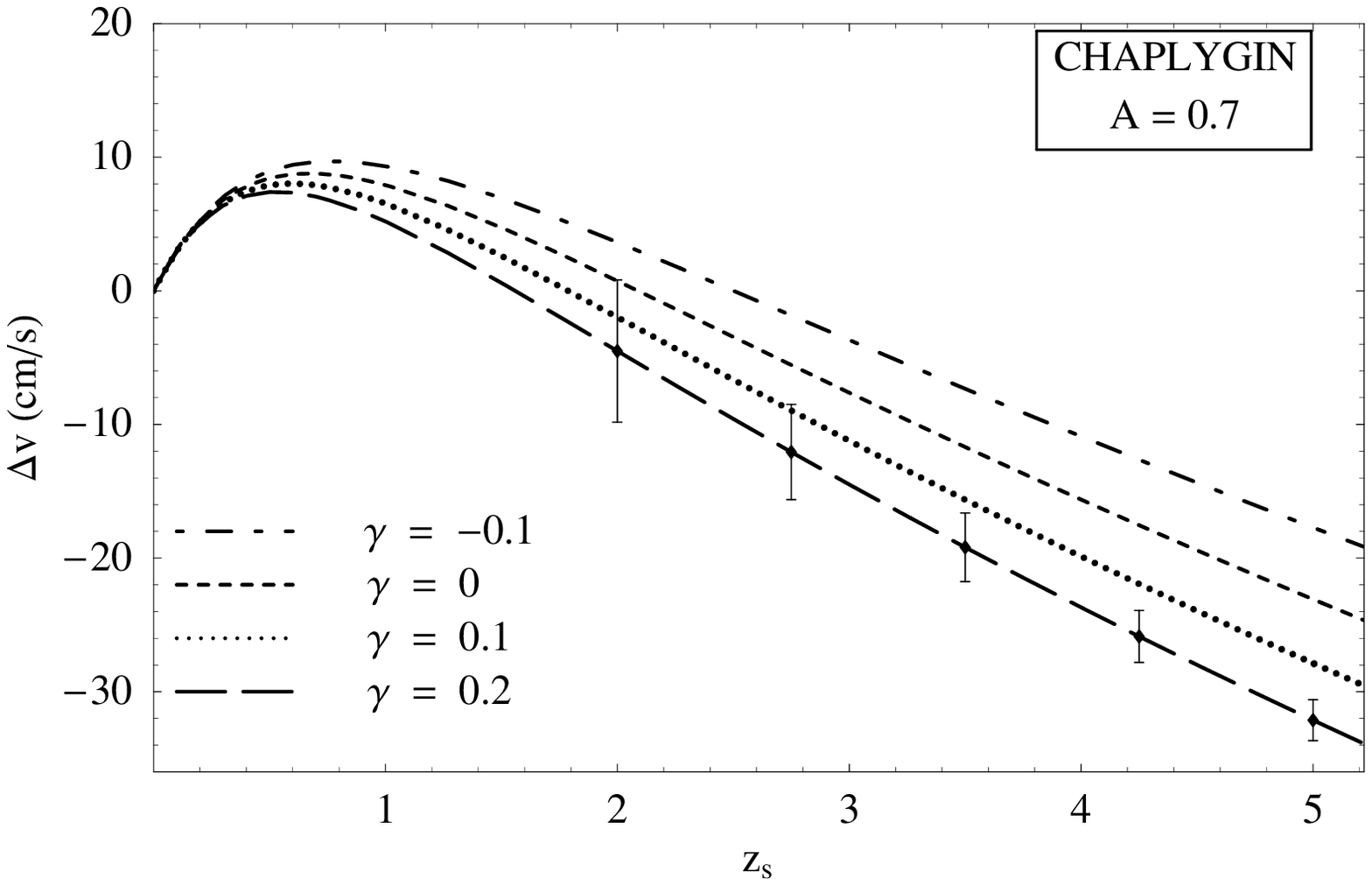}%
   \includegraphics[width=7cm]{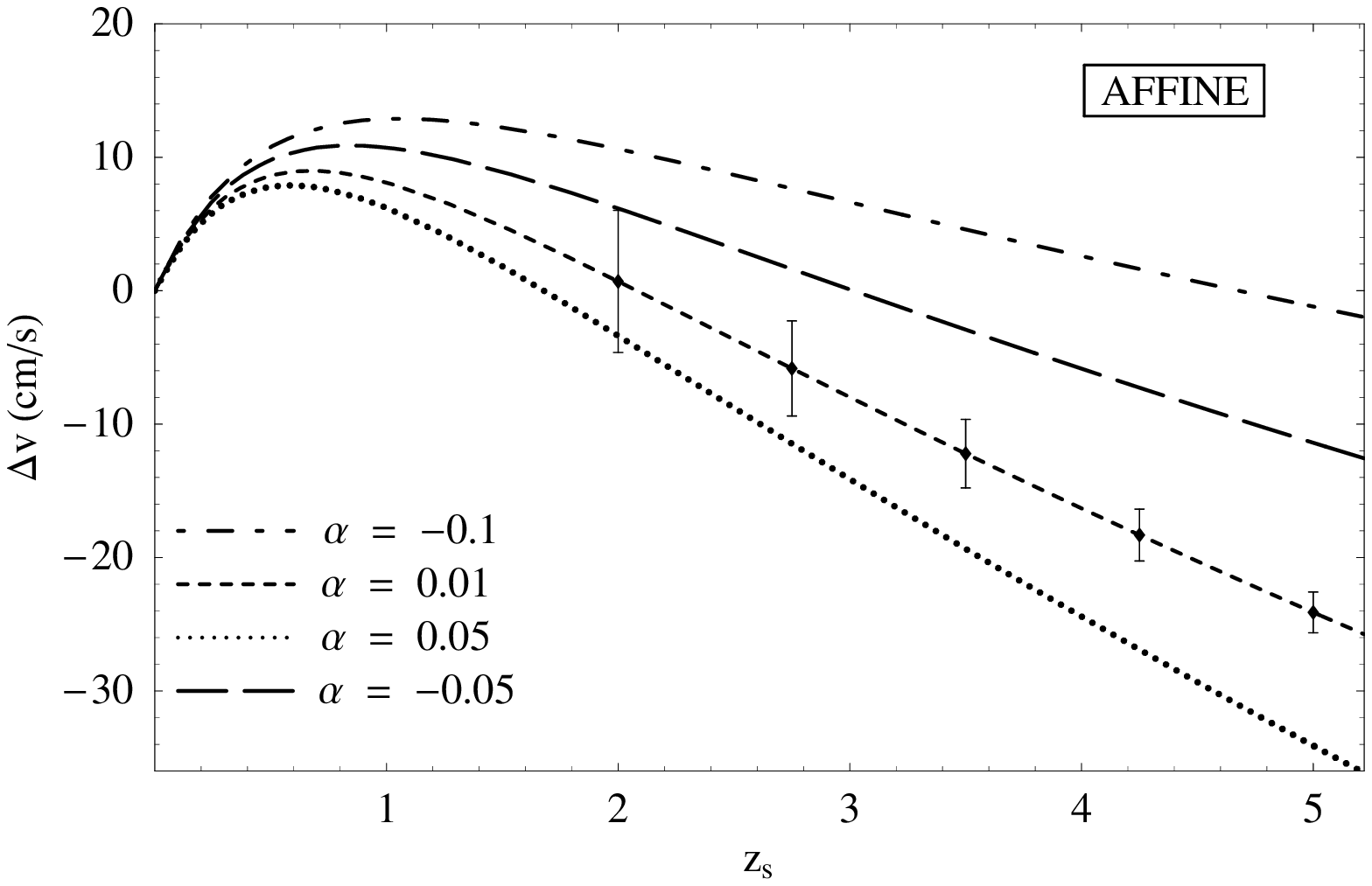}
    \caption{The apparent spectroscopic velocity shift over a period $\Delta t_0=30$ years, for a source at redshift $z_s$, for the models described in the text. From top to bottom and from left to right: the $\Lambda$CDM (Sect. \ref{lambda}), dark energy with a constant equation of state (Sect. \ref{constantw}), dark energy with varying equation of state (Sect. \ref{linder}), interacting dark energy (Sect. \ref{coupled}), DGP (Sect. \ref{dgp}), Cardassian (Sect. \ref{cardassian}), generalized Chaplygin gas (Sect. \ref{chaplygin}), dark fluid with an affine equation of state (Sect. \ref{affine}). All the other parameters are fixed at their best fit value. The decreasing error bars are due to the assumption of a uniform distribution of QSOs over the entire redshift range (see Sect.~5 for a discussion).}
\label{fig:all}
\end{figure*}

The DGP model is the one for which we obtained the tightest constraints in our analysis: $\Delta\Omega_r=0.0027$ at 1$\sigma$, assuming $\Omega_r=0.13$ as a fiducial value. This is not only due to the strong dependence of the velocity shift on $\Omega_r$ (see Fig.~\ref{fig:all}), but also to the simplicity of the model, which depends on only one parameter (in this respect, this is the simplest model, together with the standard flat $\Lambda$CDM). In general, as we already discussed, it is to be expected that models with less parameters perform better.

The issue of degeneracies is crucial when looking into the behaviour of models having more than one free parameters. We start by discussing the Chaplygin model. When both $\tilde{A}$ and $\gamma$ vary freely, no interesting constraint can be obtained observing the velocity shift with the assumed QSO data: we find $\Delta{\tilde{A}}=0.42$ and  $\Delta\gamma=1.4$ (for the fiducial values $\tilde{A}=0.7$ and $\gamma=0.2$). Fixing $\tilde{A}$, on the contrary, results in a very tight bound on $\gamma$: $\Delta\gamma=0.008$. 

We then address the interacting dark energy and the affine equation of state. As it is clear from Fig. \ref{fig:all} these models show a large variability in the redshift range we are exploring: this is to be expected, since in both models the matter-like component departs from the usual $a^{-3}$ scaling, giving a distinctive signature when one looks at higher and higher redshifts. We could then naively expect tight bounds on the parameters of these models from the observation of velocity shift. Unfortunately, again, this is not the case, due to strong degeneracies among the parameters. For example, when $\Omega_\Lambda$ is allowed to vary, for the affine equation of state we find $\Delta\alpha=0.05$ (for a fiducial value $\alpha=0.02$).   
However, if we suppose that one of the parameters is known from other observations, the constraints on the others narrow considerably. If $\Omega_\Lambda$ is known, the affine parameter $\alpha$ can be reconstructed with an error $\Delta\alpha=0.005$. If, in addition, we also know that $w=-1$ we find $\Delta\xi=0.06$ for the interacting dark energy model (for a fiducial value $\xi=3$). 

The other models do not seem to have very interesting signatures to be exploited, at least in the redshift range considered in our analysis. The worst bounds are expected for the Cardassian models, which has not a large parameter dependence in the redshift range $2\la z\la 5$ (see Fig.~\ref{fig:all}). In the case of a varying equation of state, the simple parameterization adopted in our study shows significant deviations from the fiducial model only when rather extreme values of $w_a$ (already excluded by current observations) are assumed. However, alternative parametrizations might lead to different scalings, and dark energy might dominate in the redshift window probed by Lyman-$\alpha$ forests. Models with this behaviour can in principle exist (see, for example, \citealt{2000PhRvL..85.5276D}, \citealt{1995PhRvL..75.2077F}, \citealt{2005PhRvL..95n1301C}, \citealt{2006APh....26...16L}). A recent review on the dynamics of dark energy models is \cite{Linder:2007wa}.

\begin{figure*}
\centering
\includegraphics{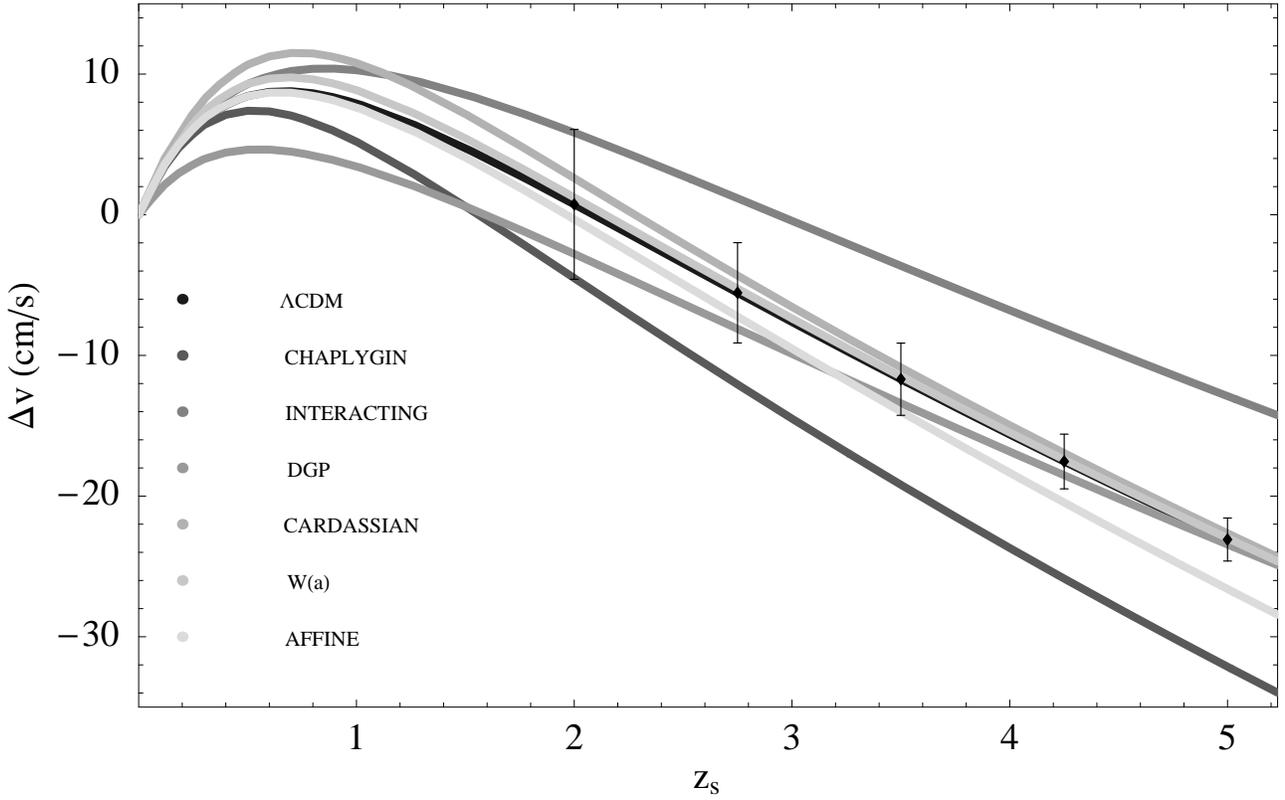}
\caption{The predicted velocity shift for the models explored in this work, compared to simulated data as expected from the CODEX experiment. The simulated data points and error bars are estimated from Eq. \ref{forecast}, assuming as a fiducial model the standard $\Lambda$CDM model. The other curves are obtained assuming, for each non-standard dark energy model, the parameters which best fit current cosmological observations.}
\label{fig:data}
\end{figure*}

A different aspect that we dealt with in our study was the potential of this cosmological tool to increase the ability to discriminate different models, when combined to other observations. Fig.~\ref{fig:data} shows the comparison among the predicted velocity shifts for all the models described earlier, assuming parameter values that are a good fit to current cosmological observations (including the peak position of the CMB anisotropy spectrum, the SNe Ia luminosity distance, and the baryon acoustic oscillations in the matter power spectrum). In other words, the models shown in Fig.~\ref{fig:data} cannot be easily discriminated using current cosmological tests of the background expansion. If we assume that the $\Lambda$CDM model is the correct one, and simulate the corresponding data points for the velocity shift, using a $\chi^2$ test we can quantify how well we can exclude the competing models based on their expected signal. As it is clear from Fig.~\ref{fig:data}, some models can be excluded with a high confidence level. In particular, we find that the Chaplygin gas model and the interacting dark energy model would be excluded at more than $99\%$ confidence level, and that the affine model would be out of the 1 $\sigma$ region. 

As a final caveat, we note that our results were obtained assuming an equal number (8) of QSOs for each of 5 redshift bins in the range $z=$2--5. (Such a uniform distribution was also assumed in the simulations performed by \citealt{2006IAUS..232..193P}.) This explains the fact that our error bars significantly decrease with redshift (see Fig.~\ref{fig:all}). Assuming a decreasing number of QSOs at higher redshifts (undoubtedly, a somewhat more realistic assumption) would result in a slight increase in the error bars for those bins. For example, assuming 3 QSOs instead of 8 in the highest redshift bin would increase the error bar for that bin of a factor 1.6. However, since the largest variability in theoretical predictions is precisely at high redhisfts, we would not expect our main conclusions to change dramatically.

\section{Conclusions}

In summary, we found that the measurement of the velocity shift with future extremely large telescopes and high-resolution spectrographs could provide interesting information on the source of cosmic acceleration, which would complement other, more traditional cosmological tools. 
From our analysis, it is also clear that the observation of velocity shift alone can be affected by strong parameter degeneracies, limiting its ability to constrain cosmological models. Contrarily to other analyses, then, our conclusion is that the uncertainties on parameter reconstruction (particularly for non-standard dark energy models with many parameters) can be rather large unless strong external priors are assumed. When combined with external inputs, however, the time evolution of redshift could discriminate among otherwise indistinguishable models. 

We also found that a reasonable time span to perform the comparison on the redshift evolution seems to be roughly three decades. Shorter time spans seem unrealistic, given the time needed to observe QSOs spectra with the necessary $S/N$. A shorter interval would also make the redshift difference too small to produce any interesting constraint on cosmological parameters.

Despite its inherent difficulties, the method has many interesting advantages. One is that it is a direct probe of the dynamics of the expansion, while other tools (e.g. those based on the luminosity distance) are essentially geometrical in nature. This could shed some light on the physical mechanism driving the acceleration. For example, even if the accuracy of future measurements will turn out to be insufficient to discriminate among specific models, this test would be still valuable as a tool to support the accelerated expansion in an independent way, or to check the dynamical behaviour of the expansion expected in general relativity compared to alternative scenarios. Furthermore, despite being observationally challenging, the method is conceptually extremely simple. For example, it does not rely on the calibration of standard candles (as it is the case of type Ia SNe) or on a standard ruler which originates from the growth of perturbations (such as the acoustic scale for the CMB) or on effects that depend on the clustering of matter (except on scales where peculiar accelerations start to play a significant role). Moreover, the errors on the measured data points decrease linearly with time and can become significantly smaller over only a few decades of observations.  Finally, it is at least conceivable that suitable sources at lower redshifts than those considered in this work could be used to monitor the velocity shift in the future. This would be extremely valuable, since some non-standard models have a stronger parameter dependence at low and intermediate redshifts (see Fig. \ref{fig:all}), that could be exploited as a discriminating tool. In \cite{1998ApJ...499L.111L} speculative possibilities of using other sources have been indicated, like masers in galactic nuclei, extragalactic pulsars or gravitationally lensed galaxy surveys: this would further extend the lever arm in redshift space and increase the ability of constraining models. Exploring the feasibility of such proposals is beyond the scope of this paper, but it may certainly be an interesting topic for further studies from observers.

\section*{Acknowledgements}
We thank Luca Pasquini for pointing out the right references to the simulated CODEX observations. We are also grateful to an anonymous referee for comments and suggestions that greatly improved the final manuscript.


\label{lastpage}

\end{document}